\documentclass[
  aip,
  amsmath,amssymb,
  reprint,
]{revtex4-1}

\usepackage{graphicx}
\usepackage{amsmath, amssymb}
\usepackage{etoolbox}
\usepackage{hyperref}

\graphicspath{{figures/}}

\makeatletter
\def\@email#1#2{%
  \endgroup
  \patchcmd{\titleblock@produce}
  {\frontmatter@RRAPformat}
  {\frontmatter@RRAPformat{\produce@RRAP{*#1\href{mailto:#2}{#2}}}\frontmatter@RRAPformat}
}

\newcommand{\FigureSetupandPrinciples}{%
\begin{figure}[t]
\centering
\includegraphics[width=\linewidth,keepaspectratio]{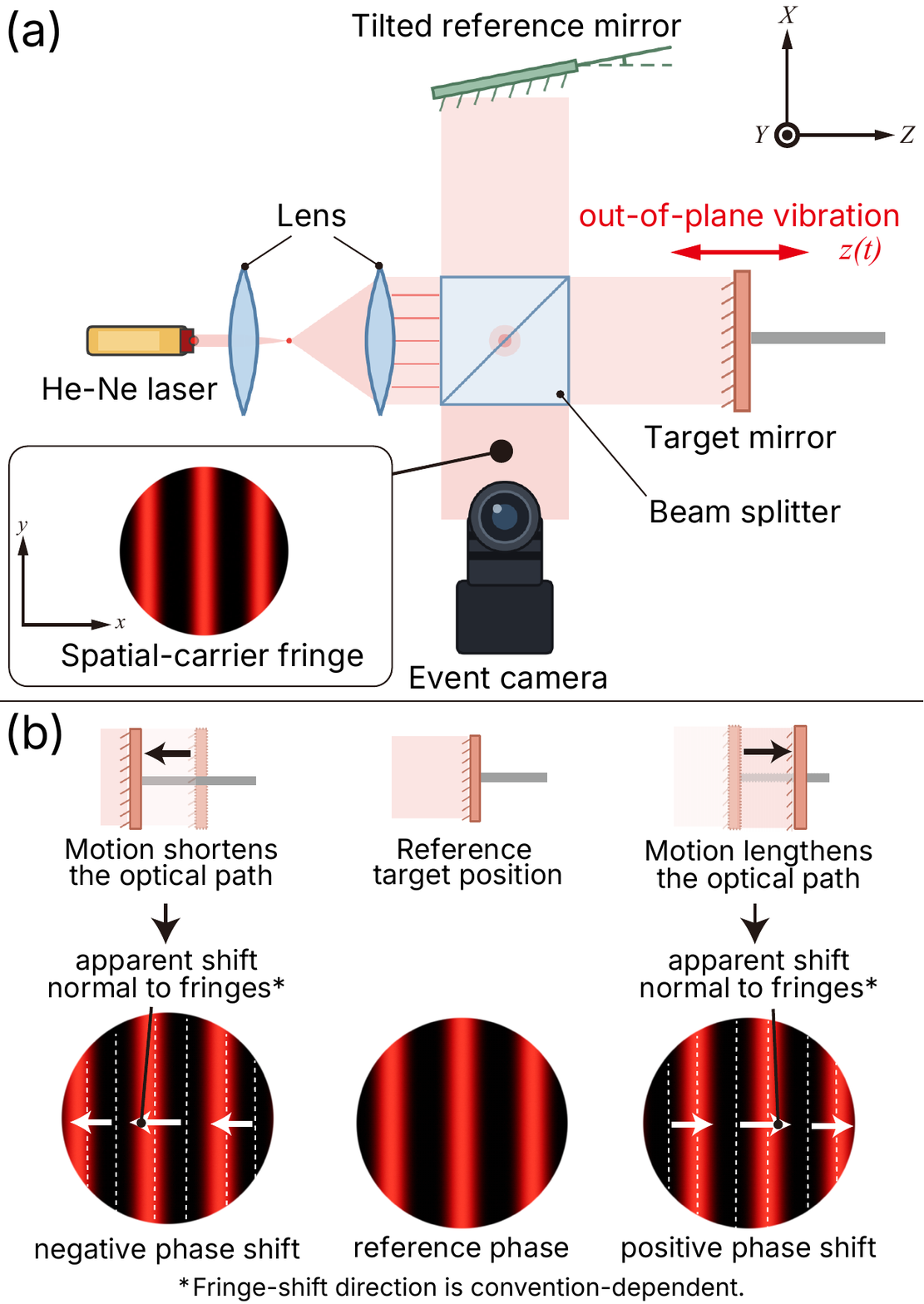}
\caption{
(a)~Experimental setup: a spatial-carrier Michelson interferometer with an event camera. A two-lens beam expander enlarges the He--Ne beam to illuminate the entire target mirror. The tilted reference mirror forms spatial-carrier fringes, while the target mirror vibrates along the optical axis.
(b)~Lateral fringe motion induced by out-of-plane vibration.}
\label{fig:setup_and_principles}
\end{figure}
}

\newcommand{\FigureObsercationOfMovingFringes}{%
\begin{figure}[t]
\centering
\includegraphics[width=\linewidth]{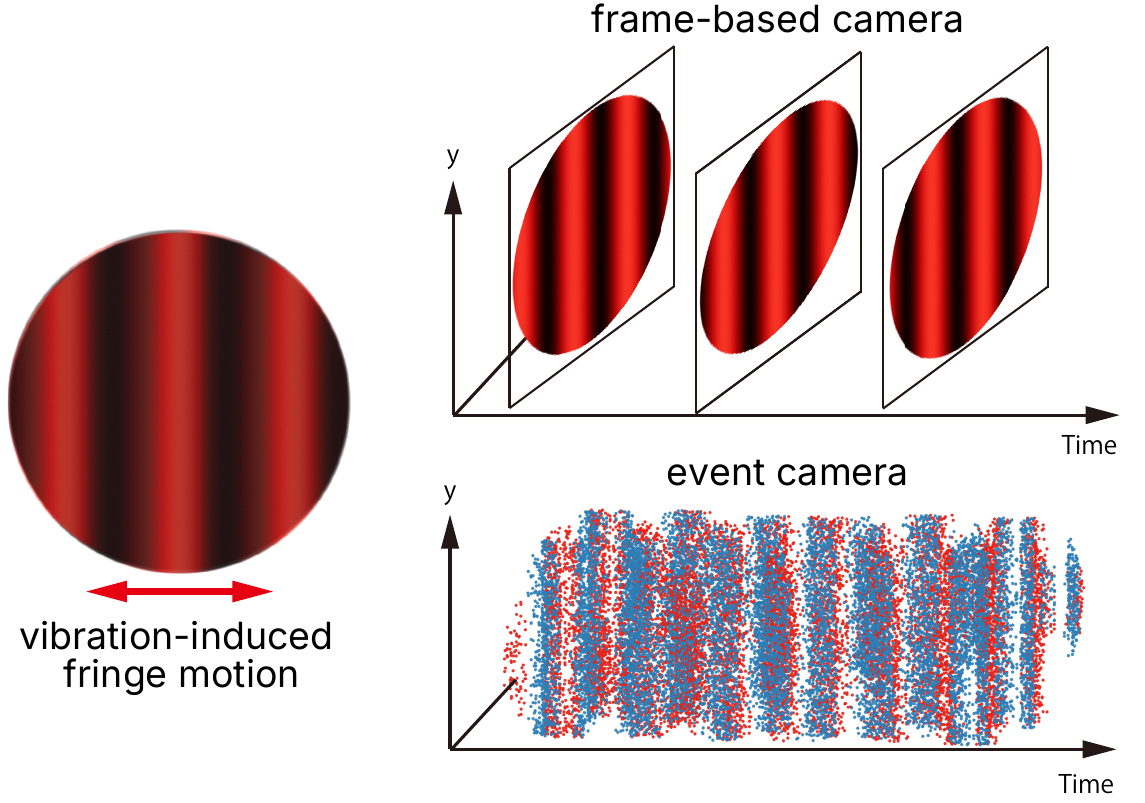}
\caption{
Frame- and event-based sampling. 
A frame camera samples discrete $x$-$y$-$t$ slices, whereas an event camera records positive (red) and negative (blue) events asynchronously in $x$-$y$-$t$ space.
}
\label{fig:obsercation_of_moving_fringes}
\end{figure}
}

\newcommand{\FigureMethod}{%
\begin{figure*}[t]
\centering
\includegraphics[width=\textwidth]{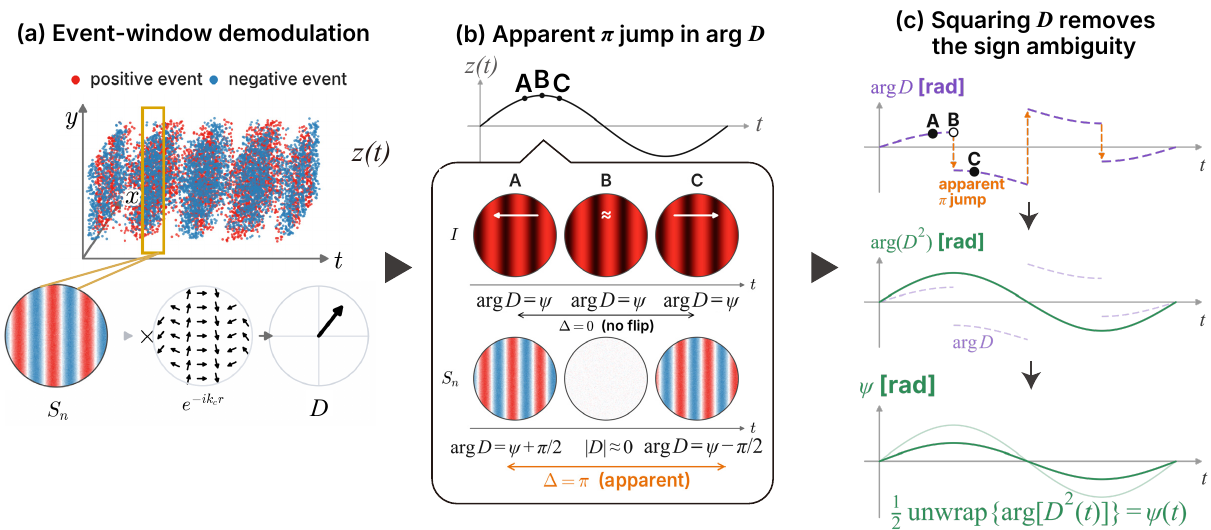}
\caption{Event-based spatial-carrier demodulation and phase recovery. (a)~Spatial-carrier demodulation of a signed event-density map. (b)~Apparent $\pi$ jump in $\arg D$ caused by the sign reversal of $\dot{\psi}(t)$ at a vibration turning point. (c)~$D^2$-based phase recovery, which removes the motion-direction sign ambiguity and enables continuous phase unwrapping.
An explanatory video of (a)--(c) is provided in the supplementary material.}
\label{fig:method}
\end{figure*}
}

\newcommand{\FigureOperatingEnvelope}{%
\begin{figure*}[t]
\centering
\includegraphics[width=\textwidth]{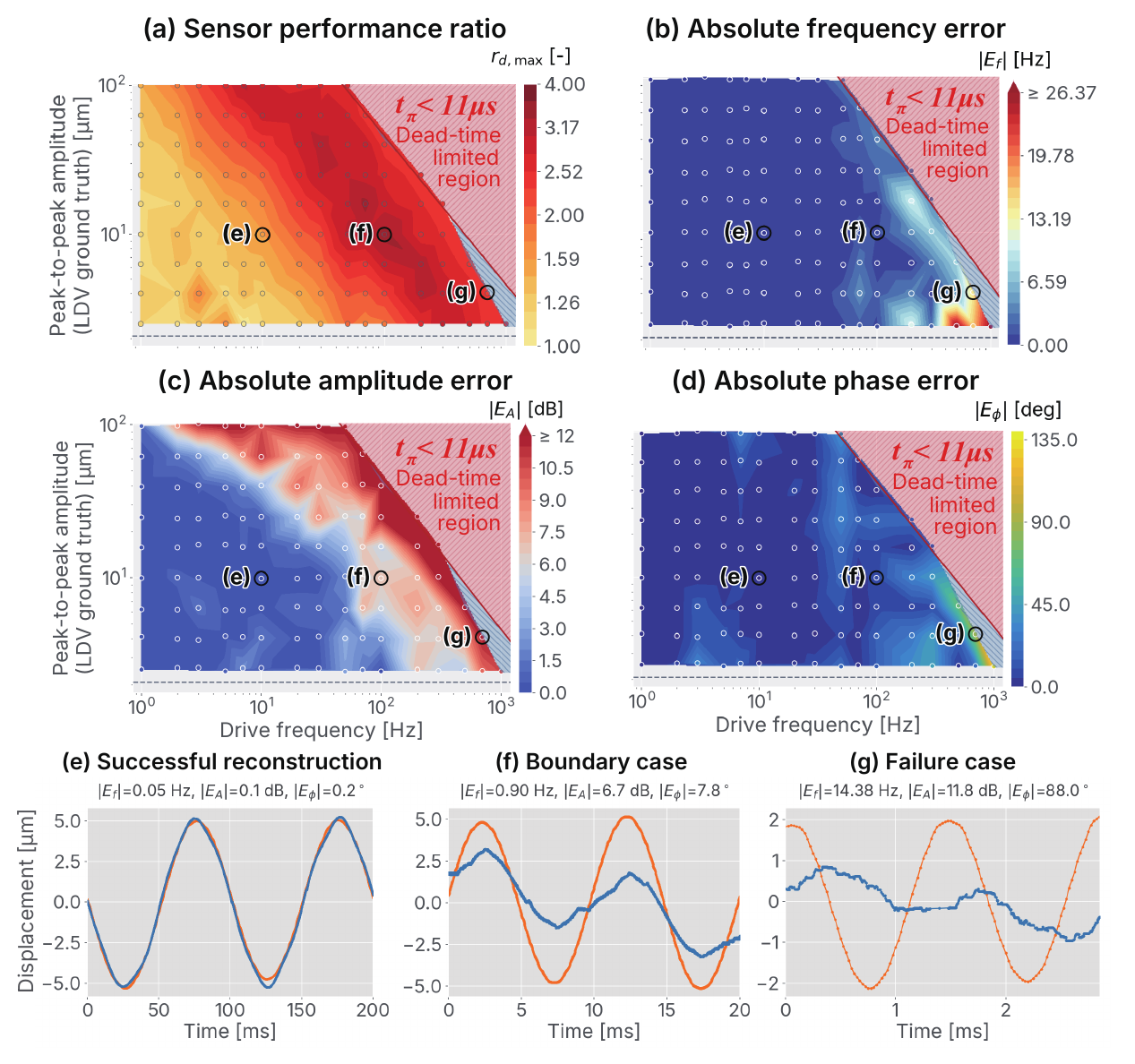}
\caption{
Operating envelope and representative reconstructions.
Top: maps of (a)~sensor-performance ratio $r_{d,\max}$,
(b)~absolute frequency error $|E_f|$,
(c)~amplitude error $|E_A|$, and
(d)~absolute phase error $|E_\phi|$ over drive frequency $f$ and
LDV-measured peak-to-peak displacement $W_{\mathrm{pp}}$.
Blue hatching marks shaker-voltage-limited, unmeasured conditions.
Color scales are clipped at 26.37\,Hz for $|E_f|$ and 12\,dB for $|E_A|$; values above these limits are shown in the darkest red.
Bottom: LDV and event-based waveforms for representative
(e)~successful, (f)~boundary, and (g)~failure cases at $(f,W_{\mathrm{pp}})=(10~\mathrm{Hz},9.9~\mu\mathrm{m})$, $(100~\mathrm{Hz},9.9~\mu\mathrm{m})$, and $(700~\mathrm{Hz},4.1~\mu\mathrm{m})$, respectively.
Orange and blue curves denote LDV and the proposed reconstruction, respectively.
}
\label{fig:envelope}
\end{figure*}
}

\newcommand{\FigurePatchReconstruction}{%
\begin{figure*}[t]
\centering
\includegraphics[width=0.96\textwidth]{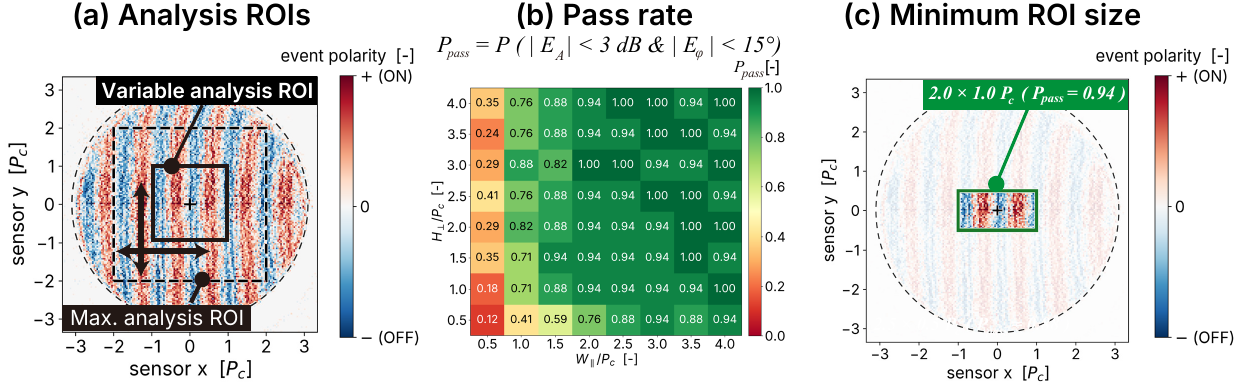}
\caption{Local-ROI reconstruction performance.
(a)~Measured signed event map of the spatial-carrier fringe events within the circular field of view, overlaid with the maximum analysis ROI and an example variable analysis ROI.
(b)~Pass fraction $P_{\mathrm{pass}}$ versus ROI width $W_{\parallel}/P_c$ and height $H_{\perp}/P_c$, with the value annotated in each cell.
(c)~The same map as (a), overlaid with the minimum ROI size: the $2.0\times1.0\,P_c$ ROI ($P_{\mathrm{pass}}\ge0.94$); the map outside this ROI is dimmed.
}
\label{fig:local_roi}
\end{figure*}
}

\makeatother
\begin{document}

\title[Event-Based Spatial-Carrier Interferometry for Surface-Normal Vibration-Waveform Reconstruction]{Event-Based Spatial-Carrier Interferometry for Surface-Normal Vibration-Waveform Reconstruction}

\author{Ryogo Niwa}
\affiliation{School of Informatics, College of Media Arts, Science and Technology, University of Tsukuba, Kasuga Campus Kasuga 1-2, Tsukuba, Ibaraki, 305-8550, Japan}
\affiliation{R\&D Center for Digital Nature, University of Tsukuba, Tsukuba, 305-8550, Ibaraki, Japan}

\author{Tatsuki Fushimi}
\affiliation{R\&D Center for Digital Nature, University of Tsukuba, Tsukuba, 305-8550, Ibaraki, Japan}
\affiliation{Institute of Library, Information and Media Science, University of Tsukuba, Tsukuba, 305-8550, Ibaraki, Japan}
\affiliation{Tsukuba Institute for Advanced Research (TIAR), University of Tsukuba, 1-1-1 Tennodai, Tsukuba, 305-8577, Ibaraki, Japan}

\author{Yoichi Ochiai}
\affiliation{R\&D Center for Digital Nature, University of Tsukuba, Tsukuba, 305-8550, Ibaraki, Japan}
\affiliation{Institute of Library, Information and Media Science, University of Tsukuba, Tsukuba, 305-8550, Ibaraki, Japan}
\affiliation{Tsukuba Institute for Advanced Research (TIAR), University of Tsukuba, 1-1-1 Tennodai, Tsukuba, 305-8577, Ibaraki, Japan}
\affiliation{Pixie Dust Technologies, Inc., Chuo-ku, 104-0028, Tokyo, Japan}

\email{niwa.ryogo@digitalnature.slis.tsukuba.ac.jp}

\date{\today}

\begin{abstract}
Non-contact measurement of small vibrations perpendicular to a surface supports the evaluation of mechanical structures, but in camera-based interferometry, increasing the frame rate makes a trade-off with the field of view and spatial resolution.
By recording only brightness changes, event cameras avoid this trade-off and reach high temporal and spatial resolution; our previously reported event topology-based visual vibrometer recovers vibration from apparent motion.
This high-speed, high-resolution sensing is well suited to full-field measurement, yet such vibration produces too little apparent motion to capture its waveform.
Here we show that event-based spatial-carrier interferometry reconstructs that waveform from moving interference fringes.
That displacement moves the fringes, and signed event-density maps built from the event stream are demodulated at the spatial carrier to recover the interferometric phase and fix the otherwise ambiguous motion direction at turning points.
Reconstructed waveforms agree with laser Doppler vibrometry over broad drive-frequency and amplitude ranges, with limits set by the maximum fringe speed and the sensor performance.
Reconstruction is limited by a minimum aperture of about two fringe periods along the carrier and one along the fringes, which allows the surface to be mapped region by region.
These results provide an empirical basis for full-field, spatially resolved interferometric vibrometry with event cameras as a non-contact measurement technique.
\end{abstract}

\maketitle

% =====================================================================
% Intro
% =====================================================================
Non-contact vibration measurements are widely used for evaluating mechanical systems and infrastructure, including aerospace structures.
Laser Doppler vibrometry (LDV)~\cite{Rothberg2017} measures single-point vibration velocity with high sensitivity.
However, LDV must scan the measurement point across the surface to obtain vibration distributions or mode shapes. 
This scanning requirement makes it difficult to capture the spatial distribution of out-of-plane vibration, that is, vibration perpendicular to the surface, across the full field of view simultaneously.

Camera-based interferometry enables full-field measurement of small out-of-plane displacements.
These methods recover an interferometric phase from fringe intensity distributions and convert phase changes into optical-path-length changes~\cite{Leendertz1970, Pedrini1994-px, Pedrini2006, Fu2007holo, Hung1973, Steinchen2003}. 
Phase-shifting methods~\cite{Bruning1974, Huntley1999-il}, spatial-carrier methods~\cite{Takeda1982, Bone1986,Moore1999-xq, Pedrini2006, Fu2007holo, Fu2014}, and single-shot methods using high-speed polarization cameras~\cite{Novak2005-np} have been widely used.
In frame-based measurements, however, increasing the frame rate makes a trade-off with the field of view, spatial resolution, and spatial sampling, which keeps fast out-of-plane vibration difficult to measure over a wide field of view~\cite{Huntley1998, Fu2014}.

Event cameras offer a potential solution for high-temporal-resolution interferometric vibration measurement without reducing spatial resolution. 
An event camera asynchronously outputs an event when the log intensity change at a pixel exceeds a threshold. 
Each event contains the pixel position, timestamp, and polarity, which indicates whether the log intensity increased or decreased~\cite{Mahowald1991-retina, Lichtsteiner2008, Gallego2022}. 
Event cameras have been used for vibration measurement based on apparent object motion~\cite{Na2023-cp, Baldini2024-dd, Baldini2026-mssp, Niwa2023-cvpr, Zhou2025-iccv, Niwa2026-apl}, speckle or specular-reflection fluctuations~\cite{Howard2023-icassp, Howard2025-tpami}, and interferometric measurements such as digital holography and coherence scanning interferometry~\cite{Schober2021-oy, Uchiyama2025-xn}.
However, these methods have not combined interferometric sensitivity with out-of-plane waveform reconstruction.

Here, we propose event-based spatial-carrier interferometry as an approach to high-temporal-resolution interferometric vibrometry that addresses the frame-rate--spatial-resolution trade-off of frame-based cameras. 
We reconstruct out-of-plane vibration waveforms from asynchronous fringe-motion events by demodulating signed event-density maps and recovering the interferometric phase while resolving the motion-direction ambiguity at vibration turning points.
Comparisons with independent LDV reference measurements map the operating envelope associated with the maximum fringe speed and sensor performance, while local-ROI tests empirically estimate the carrier/event support needed for spatially resolved reconstruction.

\FigureSetupandPrinciples

%  =====================================================================
% Method
% =====================================================================
The reconstruction starts from the physical relation between out-of-plane displacement and spatial-carrier fringe motion on the event-camera sensor.
A small tilt of the reference mirror introduces this spatial carrier in the Michelson interferometer, as shown in~\autoref{fig:setup_and_principles} (a).
In the present analysis, we consider a piston-like target whose out-of-plane displacement is spatially uniform within the analysis region of interest (ROI).
In this case, the displacement is represented by a single temporal displacement $z(t)$ and a spatially uniform interferometric phase change $\psi(t)$.
As illustrated in~\autoref{fig:setup_and_principles}(b), out-of-plane motion of the target mirror changes the optical-path-length difference of the interferometer, and this phase change translates the spatial-carrier fringes along the carrier direction, namely normal to the stripe orientation.
The corresponding spatial-carrier interference intensity on the detector is
\begin{equation}
\label{equation:interference_intensity}
I(\mathbf{r},t)=a(\mathbf{r})+b(\mathbf{r})\cos\left(\mathbf{k}_c\cdot\mathbf{r}+\phi(\mathbf{r})+\psi(t)\right),
\end{equation}
where $\mathbf{r}=(x,y)$ denotes a position in the detector plane defined in~\autoref{fig:setup_and_principles}, $a(\mathbf{r})$ is the background intensity, $b(\mathbf{r})$ is the fringe contrast, $\mathbf{k}_c$ is the spatial-carrier wave vector, and $\phi(\mathbf{r})$ is a static spatial phase.
In a Michelson interferometer, the out-of-plane displacement and the interferometric phase are related by~\cite{Hariharan2007-basics}
\begin{equation}
\label{equation:psi_and_displacement}
\psi(t)=\frac{4\pi}{\lambda}z(t).
\end{equation}
Thus, the apparent fringe motion provides a direct observable of $\psi(t)$, and hence of the out-of-plane displacement $z(t)$.

As schematically shown in~\autoref{fig:obsercation_of_moving_fringes}, the event stream generated by the fringe motion is represented as
\begin{equation}
e_i=(x_i,y_i,t_i,p_i),
\end{equation}
where $p_i\in\{+1,-1\}$ denotes the polarity of the logarithmic-intensity change.
We use these polarities directly as signed measurements of local fringe-motion-induced brightness changes.

\FigureObsercationOfMovingFringes

The reconstruction problem can therefore be formulated as estimating the temporal interferometric phase $\psi(t)$ from the event stream.
As outlined in~\autoref{fig:method}, we first convert the event stream into signed event-density maps and then demodulate their spatial-carrier component to obtain a complex signal containing $\psi(t)$.
After recovering $\psi(t)$ from this signal, the displacement waveform follows directly from~\autoref{equation:psi_and_displacement}.

\FigureMethod

We first converted the event stream into signed event-density maps, forming each map from a fixed number $N$ of consecutive events.
Because the event rate scales with fringe speed, fixing the event count makes the temporal span of each map adapt to the motion, shortening it during rapid motion and lengthening it near turning points.
Thus, $N$ sets the event support per map and the balance between temporal resolution and demodulation stability.
We index the maps by $n$.
For the $n$th map, the event polarities were accumulated on the sensor grid and Gaussian-smoothed with a standard deviation of $\sigma=1.5$ to suppress pixel-scale sparsity and local fluctuations.
The resulting smoothed signed event-density map is denoted by $S_n(x,y)$.
The representative time of the $n$th map was assigned as the midpoint between the boundary-event timestamps, $t_n=(t_{n,\mathrm{start}}+t_{n,\mathrm{end}})/2$.

For spatial-carrier demodulation, the carrier wave vector $\mathbf{k}_c$ was estimated once for each event record.
For a stable carrier estimate, we averaged the 2D FFT power spectra of 64 Hann-windowed, unsmoothed event maps evenly spaced over the record, and, after excluding the DC neighborhood, took $\mathbf{k}_c=(k_x,k_y)$ from the carrier peak.

Using this carrier estimate, we demodulated the spatial-carrier component of each $S_n$, whose complex envelope carries the desired interferometric phase, as illustrated in~\autoref{fig:method}(a).
Multiplying by the complex conjugate carrier, $\exp[-i(k_xx+k_yy)]$, shifts this phase-bearing carrier component toward zero spatial frequency.
The subsequent low-pass filtering extracts the shifted baseband component while rejecting the conjugate carrier component, residual non-carrier components, and high-spatial-frequency noise.
The complex demodulated field is computed as
\begin{equation}
D_n(x,y)=\mathrm{LPF}\!\left\{S_n(x,y)\exp\!\left[-i(k_xx+k_yy)\right]\right\},
\end{equation}
where $\mathrm{LPF}$ denotes low-pass filtering.
Because the recovered phase is spatially uniform for piston-like motion, we sampled $D_n(x,y)$ at the ROI center to minimize filtering boundary effects and obtain $D(t_n)$.

The demodulated complex time series \(D(t)\) carries the desired phase \(\psi(t)\), but it also contains a direction-dependent sign associated with the phase velocity \(\dot{\psi}(t)\); it is therefore not simply proportional to \(\exp[i\psi(t)]\).
Because events record signed temporal changes in log intensity, the signed event-density map is approximately proportional to \(\partial_t \log I(\mathbf{r},t)\).
For the fringe pattern in Equation~(1), this temporal derivative introduces a factor proportional to \(\dot{\psi}(t)\).
Thus, up to a constant complex factor, the demodulated signal can be written as
\begin{equation}
D(t)\propto \dot{\psi}(t)\exp[i\psi(t)] .
\end{equation}
When the vibration reverses direction at a turning point, \(\dot{\psi}(t)\) changes sign.
This sign change appears as an apparent \(\pi\) jump in \(\arg D(t)\), as shown in~\autoref{fig:method}(b).
To remove this direction-dependent sign, we square the signal:
\begin{equation}
D^2(t)\propto \dot{\psi}^2(t)\exp[i2\psi(t)] .
\end{equation}
Squaring makes the velocity factor nonnegative, allowing $\psi(t)$ to be recovered as one half of the unwrapped phase of $D^2(t)$, up to an arbitrary constant phase offset.
The displacement waveform was then obtained from~\autoref{equation:psi_and_displacement}.
The absolute displacement sign depends on the carrier-vector and optical-phase conventions.
We therefore calibrated this sign once by matching the initial motion direction to the LDV reference waveform and then used the same sign convention for all records.

% =====================================================================
% Experimental setup and evaluation
% =====================================================================

We first evaluated the method using an empirically selected $360\times360$~px analysis region covering the entire illuminated mirror rather than the full sensor.
This experiment assessed waveform-reconstruction accuracy and mapped the drive-frequency--displacement-amplitude operating envelope.
The setup used the spatial-carrier Michelson interferometer shown in~\autoref{fig:setup_and_principles} (a).
A He--Ne laser ($\lambda=632.8$~nm) and an EVK3 event camera were used. 
A two-lens beam expander before the beam splitter enlarged the beam, keeping it approximately collimated, to illuminate the entire target mirror.
A 20-mm-diameter, 5-mm-thick mirror bonded to a shaker was driven sinusoidally along the optical axis.
A scanning LDV, Polytec-500-3D-HV-Xtra, confirmed piston-like motion in the observed region.
The carrier pitch was not optimized and yielded about six periods across the selected analysis region (\(P_c \approx 51~\mathrm{px}\)) on average.
For this analysis, reconstruction used fixed-event-count windows of $N=500$ events, which a sensitivity analysis supported as a practical balance between sparse maps and temporal averaging.

Each drive condition was specified by the frequency $f$ and the LDV-measured peak-to-peak displacement $W_{\mathrm{pp}}$; the tested conditions covered the accessible \((f, W_{\mathrm{pp}})\) plane broadly and roughly uniformly on logarithmic axes.
The drive start time of each record was determined from an external trigger. The middle 60\% of the driven interval was treated as the steady-state interval, and the first two cycles in that interval were analyzed. 
Simultaneous LDV and interferometric event-camera measurements were not possible because both require near-normal optical access and simultaneous operation can disturb the fringe-event stream.
The LDV reference was instead measured under the same drive conditions with the shared trigger alone, which makes the frequency, amplitude, and phase of the two measurements directly comparable.
The integrated LDV displacement was used as the reference waveform for the evaluation below.

\FigureOperatingEnvelope

To evaluate the reconstructed displacement, we used three metrics: the frequency, amplitude, and phase errors.
The frequency error verifies, independently of the spectral-peak estimate used for the other two metrics, that the reconstructed waveform indeed oscillates at the drive frequency.
Each frequency was estimated as the median, over the central cycle of the same two-cycle window, of the ridge frequency of a Morlet continuous wavelet transform ($\omega_0=6$), searched within 0.6--1.4 times the drive frequency to exclude harmonics and low-frequency environmental components.
The frequency error was defined as
\begin{equation}
E_f=f_{\mathrm{event}}-f_{\mathrm{LDV}}.
\end{equation}
Each waveform was then linearly interpolated onto a uniform time grid, and the largest non-dc peak of its FFT was taken as the fundamental. The amplitude and phase were read from the magnitude and argument of this peak, without assuming the drive frequency.
The amplitude error was defined as
\begin{equation}
E_A=20\log_{10}\left(\frac{A_{\mathrm{event}}}{A_{\mathrm{LDV}}}\right),
\end{equation}
where $A_{\mathrm{event}}$ and $A_{\mathrm{LDV}}$ denote the fundamental amplitudes of the event-based and LDV waveforms.
The phase error was defined as
\begin{equation}
E_\phi
=
\operatorname{Arg}
\left(
e^{j(\phi_{\mathrm{event}}-\phi_{\mathrm{LDV}})}
\right).
\end{equation}
For the operating-envelope analysis, we used the absolute errors $|E_f|$, $|E_A|$, and $|E_\phi|$, the latter in degrees.

To interpret the operating envelope in terms of the sensor-side recording limits of the event camera, namely its per-pixel dead time and readout bandwidth, we characterized each event record using two diagnostics: the shortest time \(t_\pi\) in which the phase advances by $\pi$, and the sensor-performance ratio \(r_d\).

When the interferometric phase is sampled at discrete times, it can be tracked correctly only if the change between successive samples stays below $\pi$, the Nyquist condition for phase sampling~\cite{Huntley1993-tpu}.
This limit corresponds to a fringe shift of half the fringe period; larger shifts make the fringe-motion direction ambiguous.
For an event camera, the sampling interval of each pixel is bounded below by the dead time $\tau_{\mathrm{dead}}=11~\mu\mathrm{s}$, during which the pixel cannot record another event~\cite{Peter2023:reflactory}.
For a sinusoidal displacement with drive frequency \(f\) and peak-to-peak displacement \(W_{\mathrm{pp}}\), the maximum interferometric phase speed is
\begin{equation}
|\dot{\psi}|_{\max}
=
\frac{4\pi^2 f W_{\mathrm{pp}}}{\lambda}.
\end{equation}
The time for the phase to advance by $\pi$ at this maximum speed is
\begin{equation}
t_\pi
=
\frac{\pi}{|\dot{\psi}|_{\max}}
=
\frac{\lambda}{4\pi f W_{\mathrm{pp}}},
\end{equation}
and thus $t_\pi<\tau_{\mathrm{dead}}$ marks a dead-time-limited regime in which the fringe motion is undersampled at every pixel.
We therefore overlaid the corresponding \(t_\pi=\tau_{\mathrm{dead}}\) boundary on the \((f, W_{\mathrm{pp}})\) maps in~\autoref{fig:envelope}(a)--(d) as a physical reference rather than a sharp success--failure threshold.
Based on this a priori sensor-side estimate, drive conditions expected to satisfy \(t_\pi<\tau_{\mathrm{dead}}\) were not targeted for the quantitative operating-envelope measurement.
A small remaining high-frequency region below this boundary was not measured because the required shaker voltage exceeded the available range.

When fast fringe motion produces events across many pixels at once, the total event rate can exceed the finite readout bandwidth of the event camera~\cite{Gallego2022, Finateu2020}, and events can then be lost.
The second diagnostic, $r_d$, tests for such bandwidth overloading through the sensor-ROI dependence of the recorded event count.
Specifically, we compared the event count recorded with a reduced hardware-windowed sensor ROI, $N_{\mathrm{hw}}(s_y)$, with the count obtained by software-cropping the same region from the full-sensor record, $N_{\mathrm{sw}}(s_y)$:
\begin{equation}
r_d(s_y)=\frac{N_{\mathrm{hw}}(s_y)}
{N_{\mathrm{sw}}(s_y)} .
\end{equation}
A smaller sensor ROI lowers the total event rate; values close to one therefore indicate weak ROI dependence, whereas $r_d>1$ indicates that fewer events were recorded from the same region during full-sensor acquisition.
We used the maximum of $r_d(s_y)$ over ROI heights $s_y$ of $1/2$, $1/3$, and $1/4$ of the full sensor height, denoted $r_{d,\max}$, as the representative sensor-performance measure.

% =====================================================================
% Results
% =====================================================================
% --- Operating envelope of event-based spatial-carrier interferometry ---
\FigurePatchReconstruction
Using the sensor-performance and waveform-evaluation metrics defined above, we mapped where reconstruction remained accurate or degraded over the explored \((f, W_{\mathrm{pp}})\) range.
Small frequency, amplitude, and phase errors were obtained over a broad part of this range, largely in the region where \(r_{d,\max}\) remained close to one in~\autoref{fig:envelope}(a), as shown in~\autoref{fig:envelope}(b)--(d).
In the representative successful case shown in~\autoref{fig:envelope}(e), the event-based waveform agreed well with the LDV reference in frequency, amplitude, and phase.

The degraded reconstruction region roughly coincided with the region of large $r_{d,\max}$ in ~\autoref{fig:envelope}(a), consistent with event loss or suppression under high total event load.
In the boundary case shown in ~\autoref{fig:envelope}(f), the reconstructed waveform preserved the vibration periodicity but misestimated the peak-to-peak displacement. This indicates that the fundamental frequency and phase were still tracked, whereas quantitative amplitude recovery had become unstable.
At still higher event rates, as in ~\autoref{fig:envelope}(g), row-wise spatiotemporal skew was observed in the event stream, as shown in Supplementary.
This skew indicates that, under high event load, the recorded stream no longer represents a two-dimensional fringe pattern at a single effective time.
This behavior is consistent with row-based address-event-representation (AER) arbitration in this sensor architecture, which is documented to serialize and delay events when many rows are simultaneously active~\cite{Gehrig2022, Lopes2026-hs}.

The tested conditions nearest to the $t_\pi=\tau_{\mathrm{dead}}$ boundary showed severe degradation, consistent with the expectation that per-pixel recording becomes unreliable when the phase advances by $\pi$ within the pixel dead time.
The $t_\pi=\tau_{\mathrm{dead}}$ boundary is thus a dead-time-based physical reference, not an empirically fitted threshold; degradation in this high-speed region can also involve event loss, suppression, and row-wise timing skew under high total event load.

% --- Patch-wise reconstruction ---
We have reconstructed a single waveform from the selected analysis region for a piston-like target with spatially uniform motion.
For spatially resolved measurement such as mode-shape analysis, we empirically examined how small a local ROI can be and still recover the waveform, under the present carrier pitch and processing conditions.
To separate recording-bandwidth limitations from insufficient-ROI-size effects, this analysis used only drive conditions with good sensor performance ($r_{d,\max}<1.3$; 17 records).
Local ROI size was parameterized by the carrier-direction width $W_{\parallel}/P_c$ and the height $H_{\perp}/P_c$ along the fringes, where $P_c\approx51~\mathrm{px}$ is the reference carrier pitch of the full analysis region, as illustrated in~\autoref{fig:local_roi}(a).
Both parameters were swept from 0.5 to 4.0 in steps of 0.5, yielding 64 ROI-size combinations.

Each local ROI was processed with the same pipeline as the full analysis region, using only its own events.
Only two adjustments were made for the reduced ROI size.
For carrier estimation, the FFT was zero-padded fourfold to resolve the carrier peak in the coarsely sampled small-ROI spectrum. 
For reconstruction, the per-pixel event density was matched to the full-region $N=500$ maps by scaling events as $N=\max(150,\,500\,A_{\mathrm{ROI}}/A_{\mathrm{full}})$, where $A_{\mathrm{ROI}}$ and $A_{\mathrm{full}}$ are the ROI and full-region pixel counts, with a floor of 150 events for stable phase demodulation.

Because the motion is spatially uniform, the single-point LDV reference remains valid for every local ROI.
Each reconstruction was compared with the LDV reference, and the pass fraction $P_{\mathrm{pass}}$ was defined as the fraction of the 17 good-integrity records satisfying both $|E_A|<3~\mathrm{dB}$ and $|E_\phi|<15^\circ$.

The map of $P_{\mathrm{pass}}$ in~\autoref{fig:local_roi}(b) shows that the ROI limit was governed by the carrier direction: ROIs 2.5 carrier periods wide reached $P_{\mathrm{pass}}=0.88$ even at a height of 0.5 periods, whereas ROIs 0.5 periods wide stayed at $P_{\mathrm{pass}}\le0.41$ at every height.
ROIs at least $2.0$ carrier periods wide and $1.0$ period high maintained $P_{\mathrm{pass}}\ge0.94$, as illustrated in~\autoref{fig:local_roi}(c).
Demodulation requires fringe periods within the aperture, whereas the direction along the fringes only needs enough events.
The minimum ROI should therefore scale with the fringe period rather than with pixel count; a finer fringe pitch would thus allow proportionally smaller ROIs.

% =====================================================================
% Conclusion
% =====================================================================
We demonstrated event-based spatial-carrier interferometry for reconstructing out-of-plane vibration waveforms from asynchronous fringe events.
The method converts fixed-event-count windows into signed event-density maps and uses spatial-carrier demodulation with \(D^2\) phase recovery to suppress turning-point ambiguity.
LDV-referenced experiments showed accurate recovery when the event stream was not strongly suppressed and sufficient carrier/event support was available.
The observed envelope was mainly governed by maximum phase speed, sensor performance, and local support, providing an empirical basis for choosing operating conditions in spatially resolved event-based interferometric vibrometry.

\section*{SUPPLEMENTARY MATERIAL}
See the supplementary material for a video illustrating the event-based demodulation and the phase evolution and recovery in \autoref{fig:method}(a)--(c), the fixed-event-count sensitivity analysis, examples of row-wise spatiotemporal skew in the event stream, and additional details of the experimental setup and camera settings.

% =====================================================================
% Acknowledgments
% =====================================================================
\begin{acknowledgments}
This work was supported by Grant-in-Aid for Scientific Research (No. 24KJ0497).
\end{acknowledgments}

\section*{Data Availability Statement}
The data that supports the findings of this study are openly available in Zenodo at [DOI] and its supplementary material.

\bibliography{references}

\end{document}